\documentclass[aps,pra,amssymb,superscriptaddress,10pt,tightenlines,twocolumn,longbibliography]{revtex4-2}

\usepackage[english]{babel}
\usepackage{graphicx}
\usepackage{dcolumn}
\usepackage{bm}
\usepackage{color}
\usepackage{subfigure}
\usepackage[ansinew]{inputenc}
\usepackage{amsfonts}
\usepackage{amsmath}
\usepackage{amssymb}
\usepackage{ulem}
\usepackage{soul}

\def\ee{\mathrm{e}}
\def\dd{\mathrm{d}}
\def\ii{\mathrm{i}}
\def\D{\mathrm{D}}

\begin{document}

\title{Distance sensing emerging from second-order interference of thermal light}

\author{Francesco V. Pepe}
\affiliation{Dipartimento Interateneo di Fisica, Universit\`a degli Studi di Bari, I-70126 Bari, Italy}
\affiliation{INFN, Sezione di Bari, I-70125 Bari, Italy}

\author{Gabriele Chilleri}
\affiliation{School of Mathematics and Physics, University of Portsmouth, Portsmouth PO1 3QL, UK}

\author{Giovanni Scala}
\affiliation{Dipartimento Interateneo di Fisica, Universit\`a degli Studi di Bari, I-70126 Bari, Italy}
\affiliation{INFN, Sezione di Bari, I-70125 Bari, Italy}
\affiliation{International Centre for Theory of Quantum Technologies (ICTQT), University of Gdansk, Wita Stwosza 63, 80-308 Gda\'nsk, Poland}
\affiliation{Faculty of Physics, University of Warsaw, 02-093 Warsaw, Poland}

\author{Danilo Triggiani}
\affiliation{School of Mathematics and Physics, University of Portsmouth, Portsmouth PO1 3QL, UK}

\author{Yoon-Ho Kim}
\affiliation{Department of Physics, Pohang University of Science and Technology (POSTECH), Pohang 37673, Korea}

\author{Vincenzo Tamma}\email{Corresponding author: vincenzo.tamma@port.ac.uk}
\affiliation{School of Mathematics and Physics, University of Portsmouth, Portsmouth PO1 3QL, UK}
\affiliation{Institute of Cosmology and Gravitation, University of Portsmouth, Portsmouth PO1 3FX, UK}

\begin{abstract}
We introduce and describe a technique for distance sensing, based on second-order interferometry of thermal light. The method is based on measuring correlation between intensity fluctuations on two detectors, and provides estimates of the distances separating a remote mask from the source and the detector, even when such information cannot be retrieved by first-order intensity measurements. We show how the sensitivity to such distances is intimately connected to the degree of correlation of the measured interference pattern in different experimental scenarios and independently of the spectral properties of light. Remarkably, this protocol can be also used to measure the distance of remote reflective objects in the presence of turbulence. We demonstrate the emergence of new critical parameters which benchmark the degree of second order correlation, describing the counterintuitive emergence of spatial second-order interference not only in the absence of (first-order) coherence at both detectors but also when first order interference is observed at one of the two detectors.
\end{abstract}

\maketitle

Since the discovery of the Hanbury-Brown and Twiss (HBT) effect in the 1950s \cite{HBT1,HBT2}, the measurement of correlations of light intensities, leading to counterintuitive higher-order interference effects in absence of field coherence \cite{brannen1956,purcell1956}, has triggered the development
of quantum optics \cite{Glauber1963}. In particular, the correlation measurement at the heart of HBT effect has been the working tool of all entanglement-based protocols, from studies of quantum foundations \cite{Bell,Kim2000,legero2004quantum,shalm2015strong,giustina2015significant,laibacher2018symmetries,wang2018experimental,orre2019experimental,wang2018experimental,laibacher2018symmetries,rambach2018hectometer} to quantum-enhanced technologies such as quantum imaging and lithography \cite{imagingentangled,lithography1,lithography2,mchekhova,dangelo2008toward,brida2010experimental,lopaeva2013experimental,ono2013entanglement,barreto2014quantum,pepe2016correlation,dilena2018correlation}, information \cite{qinfo1,qinfo2,tammalaibacher2014,laibachertamma2015,tamma2014sampling,tamma2016multiboson,aaronson2011computational,tillmann2013experimental,brod2019photonic,tamma2016boson}, and teleportation \cite{teleportation}. Interestingly, starting from the early 2000s, many of these effects have been replicated by exploiting the correlations of chaotic light \cite{lee2002experimental,Bennink,Pearce2015,valencia2005two,Scarcelli,oppel2012superresolving,PhysRevA.90.063836,cpi_prl,pepe2017exploring,pepe2017diffraction}. Recently, novel schemes where second-order interference occurs effectively between light propagating through two pairs of paths that are mutually incoherent at first order have been proposed \cite{Tam-Sei,cassano,tamma2018} and experimentally realized \cite{dangelo,shih_cnot,ihn2017,smith2018} in both the temporal and spatial domain.

In this manuscript, we shed new light on the physics of second order interference with thermal light and show its sensitivity to distances in different experimental scenarios where spatial coherence is absent at either one or both the detectors and turbulence may occur. Compared with previous works, in which second-order correlations were exploited only for transverse distance measurements, this work is focused on the interplay between transverse coherence on the object planes and longitudinal distances, arbitrarily different from each other, of the detected objects. We will demonstrate that second-order coherence (hence, the possibility to observe nontrivial second-order interference) is crucially related to physical parameters, including a newly-defined \textit{thermal light second-order correlation length}, that involve specific combinations of distances and transverse coherence lengths. These results not only provide a deeper understanding of second order correlation of thermal light beyond (first-order) spatial coherence, but also enable us to be sensitive to arbitrary distances between an incoherent source and an object and between an object and a detector. This is the case even when first-order interference cannot provide information on such parameters. 

\begin{figure}
\centering
\subfigure[]{\includegraphics[height=0.25\textheight]{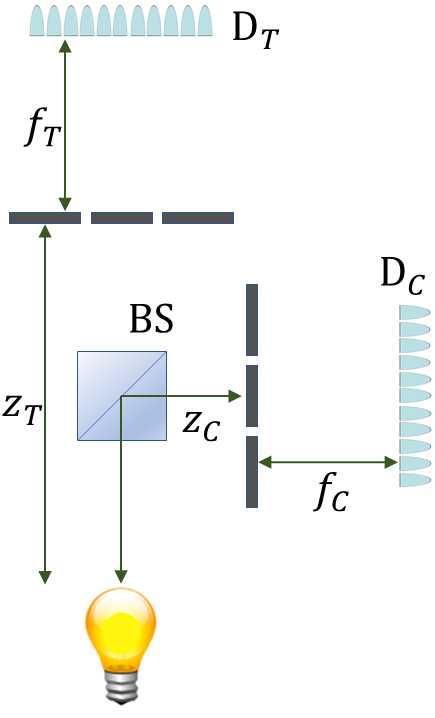}}
\hspace{0.2cm}
\subfigure[]{\includegraphics[height=0.255\textheight]{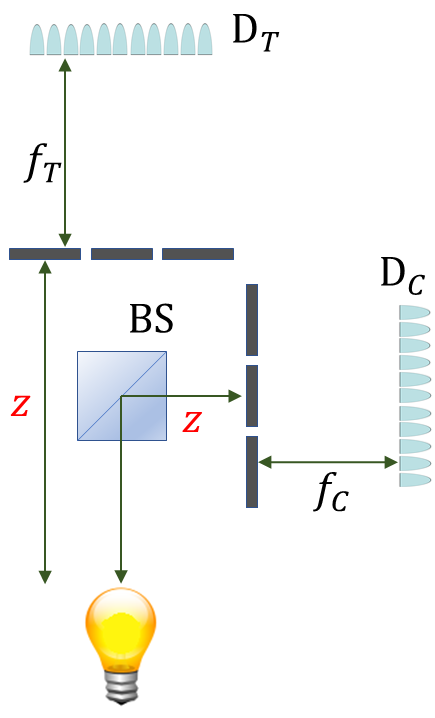}}
\subfigure[]{\includegraphics[height=0.25\textheight]{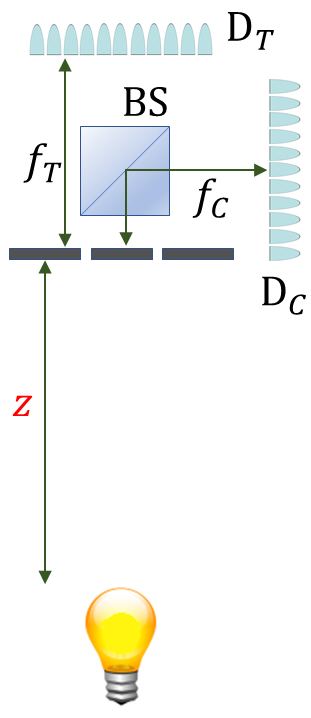}}
\hspace{0.6cm}
\subfigure[]{\includegraphics[height=0.25\textheight]{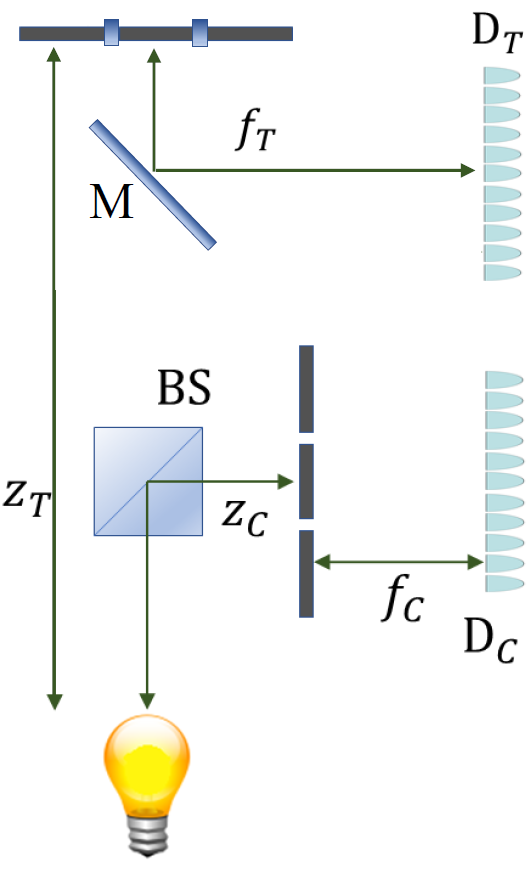}}
\caption{Panel (a): Distance sensing interferometric scheme to measure either the distance $f_T$ or $z_T$, when both the reference distances $f_C$ and $z_C$ are controlled; the source emits narrow-bandwidth thermal light; the double-slit masks are followed by two spatially resolving detectors, enabling spatial correlation measurements. Panel (b): Case of identical distances between source and masks ($z_T=z_C=z$), ideal to measure only the distance $f_T$. Panel (c): Configuration with a single mask and the beam splitter placed after the mask, equivalent in the outcome of the correlation measurements to the case (b) when the two masks are identical. Panel (d): a possible realization of the interferometric scheme in the case of a mask with reflective ``slits'', which additionally features a semi-transparent mirror placed between the beam splitter and the mask in path $T$.}\label{fig:sensing1}
\end{figure}

Our results also lay the foundations of novel protocols for distance sensing, where no frequency information about the employed thermal light is required. This can integrate and improve state-of-the-art applications, such as those based on pulsed light (e.g., time-of-flight cameras \cite{remondino2016tof}) or first-order interference (e.g., coherent LIDAR \cite{menzies1985lidar}), tasks in metrology and information processing \cite{lee2002experimental,Pearce2015,valencia2005two,oppel2012superresolving,tammalaibacher2014,PhysRevA.57.R1477}, as well as optical algorithms \cite{tamma_analogue_2015-1,tamma_analogue_2015,PRARapidFact,tamma2012prime,woelk2011factorization}. Interestingly, we show the ranging sensitivity of our second-order interference technique by employing simple double slit masks. In the more general experimental scenario, pictured in Fig.~\ref{fig:sensing1}(a), after beam splitting a thermal beam as in a standard HBT experiment, light propagates at the two output ports through two double-slit masks (usually a remote ``target'' mask $T$ and a ``controlled'' reference mask $C$ in the laboratory) before measurements of spatial correlation in the intensity fluctuations are performed at the two detectors. We demonstrate that the measured effective second order interference between the two pairs of paths through the upper and lower slits depends in a non trivial way on both the distances $(z_C,z_T)$ from the source to the two masks and to the  distances $(f_C,f_T)$ from each mask to the corresponding detector. By properly tuning the degree of second-order correlations through the values of the slit separations and  the distances $z_C$ and $f_C$, related to controlled mask, one can enhance the sensitivity of correlation measurements to the distances $z_T$ and $f_T$ of the remote target mask, even when first order interference provides no information on one of those distances or both.

We also show that when $z_T = z_C = z$ [Fig.~\ref{fig:sensing1}(b)], the same second-order interference pattern can be retrieved by the single-mask experiment in Fig.~\ref{fig:sensing1}(c), characterized by the robustness of the second-order correlated interference pattern with respect to turbulence. Finally, we remark that the described protocol can be implemented also in the case where one or both masks are reflective objects: Fig.~\ref{fig:sensing1}(d) depicts, for example, the case where the target object is a reflective mask and second order interference can be observed by introducing a semi-transparent mirror in the target path.

We shall determine the correlations of intensity fluctuations for the system in Fig.\ \ref{fig:sensing1}(a), in which the path $T$, where input thermal light is transmitted by a beam splitter, and the controlled path $C$, where light is reflected, go through two independent double-slit masks, respectively, before being detected by detectors $\mathrm{D}_T$ and $\mathrm{D}_C$ in the far field. We will consider the case in which the source emits thermal and quasi-monochromatic light, of central frequency $\omega=ck$, wavelength $\lambda=2\pi/k$, bandwidth $\Delta\omega=\tau_c^{-1}$, with $\tau_c$ the coherence time, and the two slits parallel to the vertical axis. Up to an irrelevant constant, the correlation between intensity fluctuations at the coordinate $x_C$ on the detector $\D_C$ and $x_T$ on $\D_T$ reads
\begin{align}\label{Gammapgen}
\Gamma (x_C,x_T) & = \langle \Delta I_C (x_C) \Delta I_T(x_T) \rangle \nonumber
\\ & = \left| \int \dd x_s\, \mathcal{S}(x_s) g_T(x_T,x_s) g_C^*(x_C,x_s) \right|^2 ,
\end{align}
where $\mathcal{S}$ is the intensity profile of the source and $g_{C,T}$ are the paraxial transfer functions on each path, which read
\begin{equation}\label{tf}
g_J(x_J,x_s) = \mathcal{K}(z_J) \int \dd x_o \ee^{\frac{\ii k}{2} \left[ \frac{(x_J - x_o)^2 }{f_J} + \frac{(x_o-x_s)^2}{z_J} \right] } A_J(x_o) 
\end{equation}
where $J=C,T$, $A_J$ is the transmission function of the mask placed along each path, $x_o$ is the coordinate of the mask plane, and $\mathcal{K}$ is a function independent of the transverse coordinates, including the effects of field attenuation with increasing distance (see Appendix). Here, we shall consider double-slits masks of negligible thickness $a_J$, with $J=C,T$, centered on the optical axis and characterized by the slit distances $d_C$ and $d_T$, respectively; their transmission functions will be approximated as
\begin{equation}
A_J(x_o) = a_J \left[ \delta\left(x_o+\frac{d_J}{2}\right) + \delta\left(x_o-\frac{d_J}{2}\right) \right],
\end{equation}
with $\delta(x)$ the one-dimensional Dirac delta distribution. The finite values of slit width induce low-frequency modulations of light detected in the far field of the masks, which can be safely neglected provided $k a_J x_J/f_J\ll \pi$ \cite{saleh}.

In general, the correlation of intensity fluctuations in Eq.~\eqref{Gammapgen} can be expressed as a finite Fourier series (see Appendix for derivation and full expressions of the Fourier coefficients): 
\begin{multline}\label{Gammafourier}
\Gamma\left(x_{C},x_{T}\right)= 1+ \mathrm{Re} \Biggl[ \sum_{J=C,T} F_{1J}(z_T) \exp\left( -i\frac{k d_{J}x_{J}}{f_{J}} \right)
\\  +\sum_{s=\pm}F_{2}^{\left(s\right)}(z_T) \exp\left[-i\left(s\frac{k d_{T}x_{T}}{f_{T}}+\frac{k d_{C}x_{C}}{f_{C}}\right) \right] \Biggr] ,
\end{multline}
apart from an overall constant factor given by the zero-spatial-frequency component.  Henceforth, we will assume the case in which the two
masks are centered on the respective optical axes, and a source with a Gaussian average intensity profile $\mathcal{S}(x_s) = \mathcal{S}_0 \exp ( - x_s^2 / 2\sigma^2)$, with the coherence length assuming the values
\begin{equation}\label{coharea}
\sigma_T=\frac{z_T}{k\sigma}, \quad \sigma_C=\frac{z_C}{k\sigma},
\end{equation}
at the two mask planes at distances $z_T$ and $z_C$, respectively. In this case, the Fourier coefficients in Eq.~\eqref{Gammafourier}
\begin{align}
F_{1C}(z_T) & = F_{1T}(z_T) = \frac{\cos(\alpha\beta/2)}{\cosh(\alpha/2)} \equiv F_1(z_T) ,  \label{eq:F1} \\ F_2^{(+)}(z_T) & = 1- F_2^{(-)}(z_T) = \frac{1}{1+\exp(\alpha)} , \label{eq:F2}
\end{align}
(see Appendix for derivation) depend on the product and on the absolute difference of the coherence areas in Eq.~\eqref{coharea} through two dimensionless critical parameters
\begin{equation}\label{eq:alphabeta}
\alpha= \frac{d_C d_T}{\ell_c^2}, \quad \beta= \sigma \frac{|\sigma_T-\sigma_C|}{\sigma_T \sigma_C} = k\sigma^2\frac{|z_T-z_C|}{z_T z_C} ,
\end{equation}
with the first one defined by the effective second-order correlation length 
\begin{equation}\label{eq:corrlength}
\ell_c = \sqrt{(1+\beta^2) \sigma_C \sigma_T} 
\end{equation}
at the transverse planes at distances $z_C$ and $z_T$ from the source.
In particular, for $z_T=z_C = z$, such second-order correlation length reduces to the first-order coherence length: $\ell_c=z/k\sigma$. Notice also that the Fourier coefficients in the correlation function $\Gamma$ in Eq.~\eqref{Gammafourier} depend on the slit distances only through their product, since the masks are both centered transverse with respect to the optical axis. Interestingly, the terms in Eq.~\eqref{Gammafourier} manifesting a correlation between the two masks, particularly between slits on opposite sides of the optical axis ($F_2^{(+)}$) and between slits on the same side ($F_2^{(-)}$), fully depend on the ratio $\alpha$ between such a product and the squared second-order correlation length $\ell_c$. Incidentally, notice that the paraxial approximation sets the limit of validity of the above results to the case in which the Fraunhofer conditions $k d_J^4 / (128 \min\{z_J^3,f_J^3\}) \ll \pi$ hold for both $J=C,T$ \cite{saleh}; if these conditions are violated, modulations in space of both the amplitude and the period of the interference patterns will occur, without significant changes in the physical interpretation.

\begin{figure}
\centering
\subfigure[\,$ \alpha=32.30,\, \beta=0 $]{\includegraphics[width=0.21\textwidth]{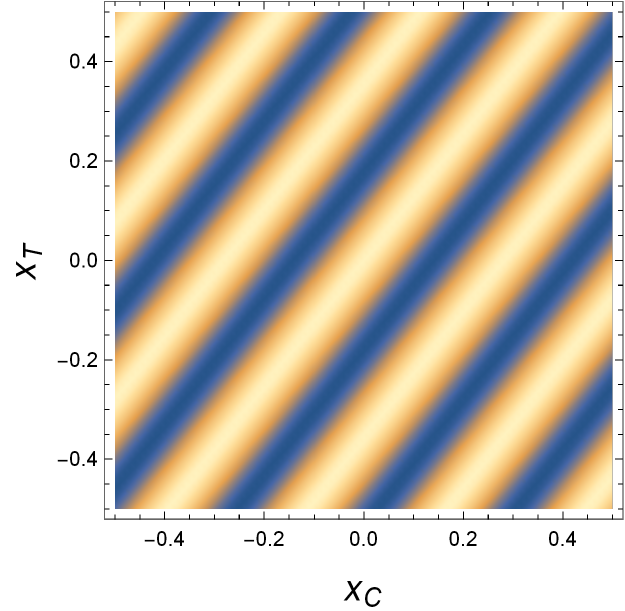}}
\subfigure[\,$\alpha=4.20,\, \beta=0.76 $]{\includegraphics[width=0.21\textwidth]{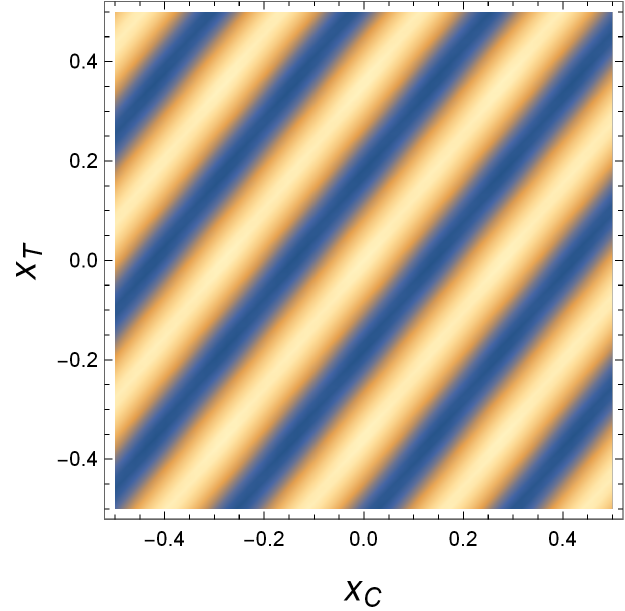}}
\subfigure[\,$\alpha=1.67,\, \beta=0.84 $]{\includegraphics[width=0.21\textwidth]{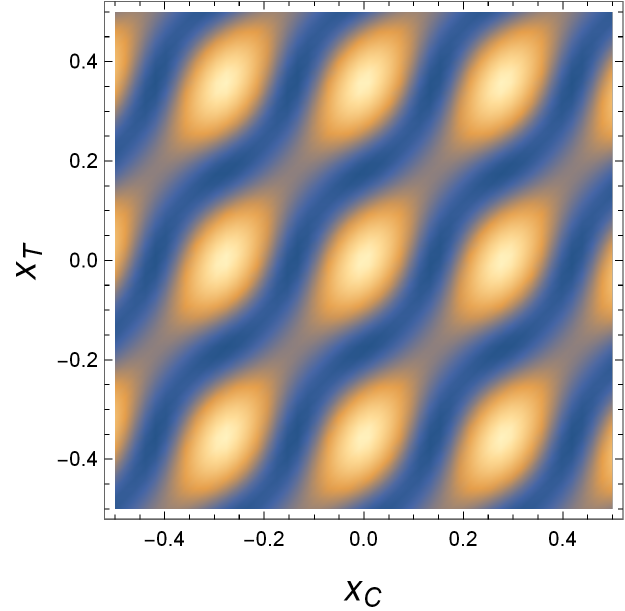}}
\subfigure[\,$\alpha=0.31,\, \beta=0.90 $]{\includegraphics[width=0.21\textwidth]{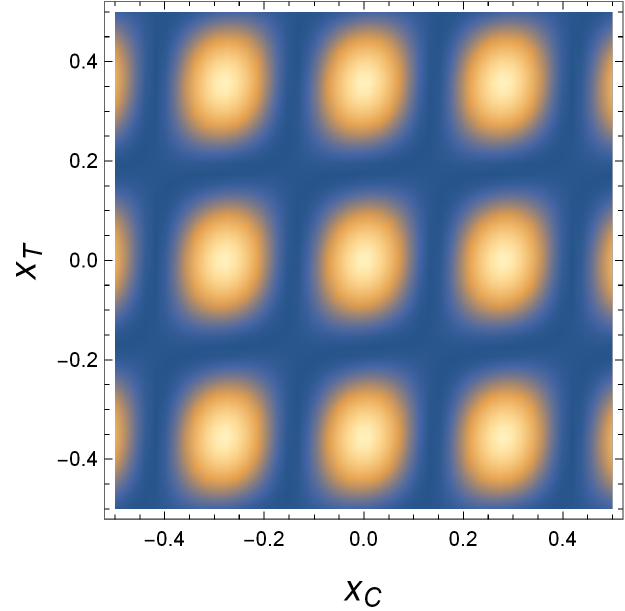}}
\caption{Density plots of the correlation function $\Gamma(x_C,x_T)$ in Eq.~\eqref{Gammafourier}, measured at different values of the target distance $z_T$ at the output of the setup in Fig.~\ref{fig:sensing1}(a), with color scale ranges from blue ($\Gamma=0$) to white ($\Gamma$ equal to its maximum). The values of the critical parameters $\alpha$ and $\beta$ in  Eq.~\eqref{eq:alphabeta} are reported in captions. The case in panel (a) is obtained for $z_T=z_C=z$, as in Fig.~\ref{fig:sensing1}(b). The second-order correlated interference pattern allows to estimate easily the length of a remote path from the detector $\mathrm{D}_T$ to the corresponding mask, as discussed in \textit{Case 2}. In panel (b), a case is shown, with $z_t/z_c=5$, in which the correlated interference pattern has still almost full visibility, despite light transmitted from the two slits of the target mask is coherent ($\sigma_T/d_T\simeq 1$). The result in panel (c) is obtained for $z_T/z_C=11.43$ and corresponds to a condition discussed in \textit{Case 3}, in which the Fourier coefficients are more sensitive to the measure of the unknown distance between one mask and the source. In panel (d), the case of a factorized interference pattern is shown, as discussed in \textit{Case 1}. The constant parameters are $\lambda=980\,\mathrm{nm}$, $d_C=0.70\,\mathrm{mm}$, $d_T=0.55\,\mathrm{mm}$, $\sigma=0.1\,\mathrm{mm}$, $z_C=70\,\mathrm{mm}$, $f_T=f_C=200\,\mathrm{mm}$.}\label{fig:Gamma}
\end{figure}

Remarkably, the measurement of the spatial frequencies in the correlation function (\ref{Gammafourier}) allows to infer the distance of the length $f_T$ of the target path from the detector $\mathrm{D}_T$ to the corresponding mask. Furthermore, an analysis of the Fourier coefficients allows us to extrapolate the value of the distance $z_T$ from the source to the target mask. In Fig.~\ref{fig:Gamma}, we show the behavior of the correlation function in the $(x_C,x_T)$ plane with varying $z_T$. We emphasize that the intensity at the detector $\mathrm{D}_T$
\begin{equation}\label{firstorder}
I(x_T) \propto 1+ \exp\left( - \frac{d_T^2}{2 \sigma_T^2} \right) \cos\left( \frac{k d_T x_T}{f_T} \right) ,
\end{equation}
is highly sensitive to $f_T$ only if $d_T\ll\sigma_T$, and to $z_T$ (through the coherence length $\sigma_T$) only if $d_T\sim\sigma_T$. Remarkably, we will now show how second-order correlation measurement allows us to retrieve this sensitivity in arbitrary ranges of $\sigma_T$ by a proper tuning of the parameters related to the mask $C$.

\begin{figure}
\centering
\subfigure[\,$\sigma=1.17$ mm]{\includegraphics[width=0.48\textwidth]{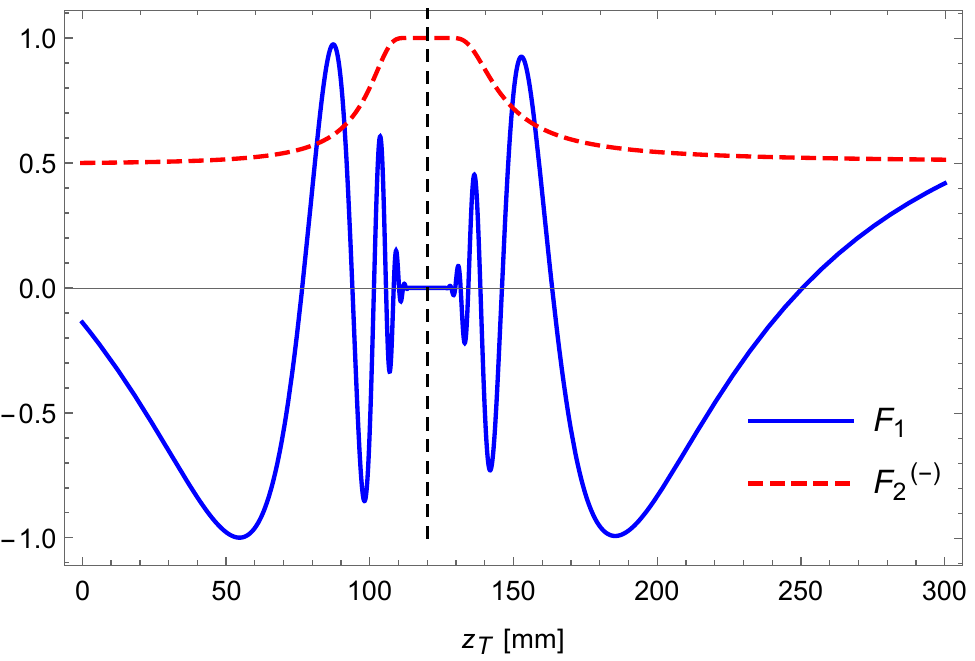}}
\subfigure[\,$\sigma=0.17$ mm]{\includegraphics[width=0.48\textwidth]{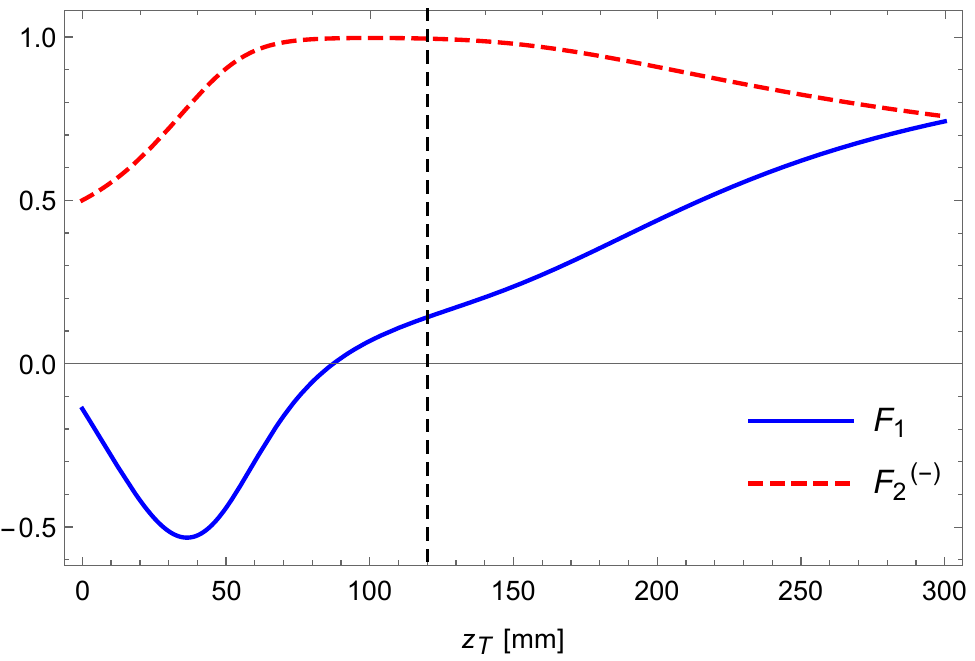}}
\caption{Plots of the Fourier coefficients $F_1$ (solid blue line) and $F_2^{(-)}$ (dashed red line), defined in Eqs.~(\ref{eq:F1})-(\ref{eq:F2}), as functions of the distance $z_T$ between source and target mask $T$, at fixed $z_C= 120\,$mm, $\lambda= 980\,$nm $d_{T}=0.08\,\mathrm{mm}$, $d_{C}=0.8\,\mathrm{mm}$, for two different values of the source width $\sigma$. The plateau around $z_C=z_T$ in panel (a) corresponds to a range in which $\alpha \gg 1$, producing correlated interference patterns analogous to the one in Fig.~\ref{fig:Gamma}(a). Notice that, in the case $z_T=z_C$ of panel (b), $\alpha\simeq 5$ is not small, providing still a non-factorized pattern despite the two slits of mask $C$ fall within the coherence area at a distance $z_C$.}\label{fig:sigma}
\end{figure}

\textit{Case 1. The factorized limit.---} We first describe the regime in which the correlation function $\Gamma(x_C,x_T)$ factorizes with respect to its two detector position variables.
By using the expressions (\ref{eq:F1})-(\ref{eq:F2}) of the Fourier coefficients, the correlation function in Eq.~\eqref{Gammafourier} factorizes as
\begin{multline}\label{eq:factorized}
\Gamma(x_C,x_T) = \Gamma_{\mathrm{f}}(x_C,x_T) \\ = 4 \cos^2\left( \frac{k d_T x_T}{2 f_T} \right) \cos^2\left( \frac{k d_C x_C}{2 f_C} \right)
\end{multline}
if and only if $F_1^2=2F_2^{(+)}=2F_2^{(-)}$, approximately occurring when $\alpha\ll 1$ and $\alpha\beta\ll 1$, as defined in Eq.~\eqref{eq:alphabeta} (the latter condition is automatically satisfied for $z_T=z_C$, where $\beta=0$). The approximate factorization entails, as a necessary condition, a bound on the product of transverse slit distances:
\begin{equation}\label{condA}
d_C d_T  \ll \ell_c^2 . 
\end{equation}
Interestingly, to satisfy the factorization conditions and hence to observe a pattern approximated by \eqref{eq:factorized}, as in Fig.~\ref{fig:Gamma}(d), it is not necessary to be in the regime where first order interference can be observed at both detectors, namely $d_C\ll\sigma_C$ and $d_T\ll\sigma_T$. Unfortunately, in such a factorized limit, the second-order correlation has a small sensitivity  to variations in $z_T$, depending on it only through its first order contribution in $\alpha$,
\begin{equation}\label{eq:distdecA}
\Gamma(x_C,x_T) - \Gamma_{\mathrm{f}}(x_C,x_T) \simeq \frac{\alpha}{2} \sin\!\left( \frac{k d_T x_T}{f_T} \right) \sin\!\left( \frac{k d_C x_C}{f_C} \right).
\end{equation}
arising from the only nonvanishing first-order corrections to the Fourier coefficients, namely $F_2^{\pm} = 1/2 \mp \alpha/4$. 
.

\textit{Case 2. The limit of second-order correlated interference pattern: higher sensitivity to $f_T$.---} Considering the form of the Fourier coefficients in Eqs.~(\ref{eq:F1})-(\ref{eq:F2}), it is evident that, when $\alpha\gg 1$, namely
\begin{equation}\label{condB}
d_C d_T  \gg \ell_c^2  ,
\end{equation}
a limit is approached in which $F_1=F_2^{(+)}=0$ and $F_2^{(-)}=1$, representing effective interference between pairs of paths associated with corresponding slits in the two masks. In this case, the intensity correlation function in Eq.~\eqref{Gammafourier} takes the form of a correlated interference pattern
\begin{equation}\label{eq:cosinesquared}
\Gamma(x_C,x_T) = \Gamma_{\mathrm{cp}} (x_C,x_T) = 2 \cos^2\left[ \frac{k}{2} \left( \frac{d_T x_T}{f_T} - \frac{d_C x_C}{f_C} \right) \right] ,
\end{equation}
The first-order corrections in $e^{-\alpha/2}$
where the orientation of the interference fringes is fixed by a specific linear combination of the two detector variables [see Fig.~\ref{fig:Gamma}(a)]. This condition is opposed to the factorized case in Eq.~(\ref{eq:factorized}) and is the most convenient in order to detect the distance $f_T$. It is sufficient to characterize the oscillation frequency along any of the directions $x_C=\mu x_T+\nu$, with $\mu\neq (d_T/d_C)(f_C/f_T)$, and $\nu$ arbitrary, in order to determine $f_T$. In particular, the frequency of the second-order interference pattern in the case $x_C=-(d_T/d_C)(f_C/f_T)x_T+\nu$ is \textit{twice} the frequency of the pattern generated at first order by coherent light impinging the double slit mask. Notice that both the directions of constant $\Gamma$ in the $(x_C,x_T)$ plane and the directions of maximal frequency are independent of the wavelength of detected light. Therefore one can also infer the distance $f_T$, for a known reference distance $f_C$, without precise knowledge of the light frequency, by determining one of these two directions or both. Knowledge of the correlation function, determined by the parameters
\begin{align}
F_1 & = 2\ee^{-\frac{\alpha}{2}}\cos(\alpha\beta)[1+O(\ee^{-\alpha})], \label{eq:equal1} \\ F_2^{(+)} & = 1- F_2^{(-)} = O(\ee^{-\alpha}) \label{eq:equal2}
\end{align}
enables one to infer (though not unambiguosly, see following discussion for Case 3) the value of $z_T$ from the amplitude of additional uncorrelated oscillations
\begin{equation}
\Gamma(x_C,x_T) - \Gamma_{\mathrm{cp}}(x_C,x_T) \simeq 2\ee^{-\frac{\alpha}{2}}\cos(\alpha\beta) \sum_{J=C,T} \cos\frac{k d_J x_J}{f_J}.
\end{equation}
However, such oscillations are exponentially suppressed with high decay value $\alpha\gg 1$ [however, notice by comparing Eqs.~\eqref{eq:equal1}-\eqref{eq:equal2} that $F_1$ is more sensitive than $F_2^{(\pm)}$ to changes in $z_T$]. Therefore, they are not so sensitive to variations in $z_T$, as reflected for example in the presence of plateaus that can occur in both $F_1$ and $F_2^{(-)}$, especially in a neighborhood of $z_T=z_C$ [see Fig.\ \ref{fig:sigma}(a)]. Interestingly, also in this case, the occurrence of full second-order correlations in Eq.~\eqref{eq:cosinesquared} for $\alpha \gg 1$ is \textit{not} necessarily determined by the absence of first-order interference at both detectors. Indeed, it is possible that the slits of \textit{one} mask fall within the coherence length on their plane, as shown in Fig.~\ref{fig:Gamma}(b). Remarkably, in the case of a single mask ($d_C=d_T=d$), represented in Fig.~\ref{fig:sensing1}(c), the condition in Eq.~\eqref{condB} for the observation of a fully correlated pattern ensures robustness to turbulence surrounding the mask, since the correlation function becomes insensitive to any local (random) phase of the field at each slit, as proved in the Appendix (see also \cite{smith2018}).

\textit{Case 3. The intermediate range: higher sensitivity to $z_T$.---} We have shown that the extremal situations considered before are not ideal in measuring the distance $z_T$. We demonstrate now how this drawback can be overcome by considering instead intermediate settings. For example, for $\beta\gtrsim\alpha^{-1}\gg 1$,
$F_2^{(\pm)}\simeq 1/2\mp \alpha/4$, as in the factorized case, while $F_1\simeq \cos(\alpha\beta/2)$ can now provide an unambiguous estimation of $z_T$ through the variation of $\alpha\beta$ in a given interval of length $\pi$. Furthermore, in the case in which $\alpha\sim 1$ and $\beta\gtrsim 1$, namely, from Eq.~\eqref{eq:alphabeta},
\begin{equation}
d_C d_T \sim \ell_c^2, \quad k\sigma^2\left|z_T - z_C \right| \gtrsim z_C z_T ,
\end{equation}
both the independent Fourier coefficients in Eqs.~\eqref{eq:F1}-\eqref{eq:F2} are sensitive to variations in the distance $z_T$. Such a condition can be observed in the plot of Fig.~\ref{fig:Gamma}(c). The measure of $F_1$ and $F_2^{(+)}$ therefore enables a combined estimate of $z_T$. As shown in Fig.~\ref{fig:sigma}, the coefficient $F_1$ is typically more sensitive than $F_2$ to small variations of $z_T$, although strongly non-monotonous, as opposed to $F_2^{(\pm)}$, with monotonicity intervals typically centered around the cosine zeros $\alpha\beta=(2n+1)\pi$, with $n\in\mathbb{N}$.
An effective strategy for the estimation of a completely unknown $z_T$ can include a two-step process, involving 1) a rougher estimate of the range of distance through the parameter $F_2$, less sensitive but characterized by only two monotonicity ranges, and 2) a more precise estimation through $F_1$.

In conclusion, we described the physics of second-order interference between a pair of double-slit masks, placed at \textit{arbitrary} distances from a common thermal source and from the detectors behind. We also demonstrate its application to sensing the distances of a remote object from the source and from the detector in different \textit{ad hoc} experimental settings; in particular, in the regime of second-order correlated interference, it provides a way to measure the distance between object and detector regardless of the spectral properties of the light.
Such a technique can be also implemented with a single mask Fig.~\ref{fig:sensing1}(c), if one is mainly interested in measuring the distance between the mask and one of the detectors. 

Our analysis sheds new light in the understanding of the emergence of \textit{second-order coherence} with thermal light and its connection to the degree of correlation of the measured second-order interference pattern. Remarkably, we have shown that the absence (presence) of second-order interference is not necessarily connected with the presence (absence) of first-order interference at \textit{both} detectors separately. On the other hand, such second-order correlations emerge from a second-order coherence which does not depend on the coherence lengths measured at the two masks independently but on a suitable combination of them in two new critical parameters,  the second order correlation length $\ell_c$ and the non-dimensional parameter $\beta$ in Eqs.~\eqref{eq:alphabeta}-\eqref{eq:corrlength}. 

These findings also provide the basis for a convenient protocol to measure the distance of reflective objects, placed either on the optical path between source and mask, or on the path between mask and detector, as shown in Fig.~\ref{fig:sensing1}(d). Such a protocol, based on the control of second-order coherence through the parameter $\ell_c$, provides the possibility to apply interference-based distance detection protocols, such as coherent LIDAR, even when first-order coherence cannot be exploited. A generalization of this analysis to masks with arbitrary transverse position with respect to the optical axis, as well as more complex spatial structures, will be addressed in future works. Our results also pave the way to interesting future research devoted to an accurate evaluation of the ultimate precision bounds of the described measurement scheme \cite{motka2016}, and the least possible error given the state of the field, quantified by the Quantum Fisher information \cite{napoli2019}.

\begin{acknowledgments}
\textit{Acknowledgments.---} This project was partially supported by the Office of Naval Research (ONR) Global (Award No.\ N62909-18-1-2153). YHK was partially supported by the National Research Foundation of Korea (Grant No.\ 2019R1A2C3004812) and the MSIT of Korea under the ITRC support program (IITP-2020-0-01606). FP and GS were partially supported by the Istituto Nazionale di Fisica Nucleare (INFN) projects PICS and QUANTUM. FP was partially supported by PON ARS 01\_00141 ``CLOSE -- Close to Earth'' of Ministero dell'Istruzione, dell'Universit\`a e della Ricerca (MIUR). GS was partially supported by the Foundation for Polish Science (IRAP project, ICTQT, contract no.\ 2018/MAB/5, co-financed by EU within Smart Growth Operational Programme) and the National Science Center (Poland) grant No.\ 2016/22/E/ST2/00559.
\end{acknowledgments}

\onecolumngrid
\appendix

\section{General form of the correlation function}

Up to an irrelevant constant, the correlation between intensity fluctuations at the coordinate $x_C$ on the detector $\D_C$ and $x_T$ on $\D_T$ reads
\begin{equation}\label{Gammapgen}
\Gamma (x_C,x_T)  = \langle \Delta I_C (x_C) \Delta I_T(x_T) \rangle  = \left| \int \dd x_s\, \mathcal{S}(x_s) g_T(x_T,x_s) g_C^*(x_C,x_s) \right|^2 ,
\end{equation}
where $\mathcal{S}$ is the intensity profile of the source and $g_{C,T}$ are the paraxial transfer functions on each path, which read
\begin{equation}\label{tf}
g_J(x_J,x_S) = -\frac{k^2}{4\pi^2 z_J f_J} \int \dd x_o \ee^{\frac{\ii k}{2} \left[ \frac{x_J^2- 2 x_o x_J}{f_J} + \frac{(x_o-x_s)}{z_J} \right] } A_J(x_o) \qquad \text{(with } J=C,T\text{)}, 
\end{equation}
where $A_J$ is the transmission function of the mask placed before each lens and $k=2\pi/\lambda$ is the photon wavenumber. Here, we shall consider the transmission function of double-slits masks of negligible thickness $a$, approximated as
\begin{equation}\label{eq:AJ}
A_J(x_o) = a \left[ \delta \left(x_o+\frac{d_J}{2}\right) + \delta \left(x_o-\frac{d_J}{2}\right) \right] .
\end{equation}
The transfer functions can be expressed as
\begin{equation}
g_J(x_J,x_S) = -\frac{a k^2}{4\pi^2 z_J f_J} \exp\left( \frac{\ii k}{2} \left( \frac{x_J^2}{f_J} + \frac{x_s^2 + d_J^2/4}{z_J} \right) \right) \sum_{q=\pm} \exp\left( -  \frac{\ii k d_J}{2} q \left( \frac{x_s}{z_J} + \frac{x_J}{f_J} \right) \right) .
\end{equation}
The above results are the basic elements to compute the correlation function \eqref{Gammapgen}. 
The expectation value on the thermal state appearing in Eq.~\eqref{Gammapgen} is practically estimated by averaging over the products of intensities measured in correspondence of a set of discrete observation times.

For general distances ($z_T\neq z_C$), the quantity in Eq.~\eqref{Gammapgen} reads
\begin{equation}\label{GammaW}
\Gamma (x_C,x_T) = \frac{a^2 k^4}{(2\pi)^4 z_C z_T f_C f_T} \Biggl| \sum_{q,q'=\pm} \mathcal{F}_{qq'}(z_C,z_T) \exp\left[ - \frac{\ii k}{2} \left( q \frac{x_T d_T}{f_T} - q'\frac{x_C x_C}{f_C} \right) \right] \Biggr|^2 ,
\end{equation}
where
\begin{equation}
\mathcal{F}_{qq'}(z_C,z_T) = \Phi \left( q\frac{d_T}{z_T} - q'\frac{d_C}{z_C} \right) \exp\left[ \frac{\ii k}{8} \left( \frac{d_T^2}{z_T} - \frac{d_C^2}{z_C} \right) \right]
\end{equation}
with
\begin{equation}\label{Phip}
\Phi(y) = \int \dd x_s \mathcal{S}(x_s) \exp \left( \frac{\ii k }{2}  \left(\frac{1}{z_T} - \frac{1}{z_C}\right)  x_s^2 - \frac{\ii k y}{2} x_s \right),
\end{equation}
coinciding with the Fourier transform of the source intensity profile in the case $z_T=z_C=z$. 

In general, the correlation of intensity fluctuations, as a function of the detector coordinates, can be expressed as a finite Fourier series: 
\begin{equation}\label{Gammafourier}
\Gamma\left(x_{C},x_{T}\right)= \frac{a^2 k^4}{(2\pi)^4 z_C z_T f_C f_T}  B \left\{ 1+ \mathrm{Re} \left[ F_{1C} e^{-i\frac{k d_{C}x_{C}}{f_{C}}}+F_{1T} e^{-i\frac{k d_{T}x_{T}}{f_{T}}}
 +\sum_{s=\pm}F_{2}^{\left(s\right)} e^{-i\left(s\frac{k d_{T}x_{T}}{f_{T}}+\frac{k d_{C}x_{C}}{f_{C}}\right)} \right] \right\} ,
\end{equation}
with
\begin{align}
B & = \sum_{q,q'=\pm} |\mathcal{F}_{qq'}(z_C,z_T)|^2 , \label{B} \\
F_{1C} & = \frac{2}{B} \left( \mathcal{F}_{--}(z_C,z_T) \mathcal{F}_{-+}^*(z_C,z_T) + \mathcal{F}_{+-}(z_C,z_T) \mathcal{F}_{++}^*(z_C,z_T) \right) , \label{F1C}  \\
F_{1T} & = \frac{2}{B} \left( \mathcal{F}_{--}^*(z_C,z_T) \mathcal{F}_{+-}(z_C,z_T) + \mathcal{F}_{-+}^*(z_C,z_T) \mathcal{F}_{++}(z_C,z_T) \right) , \label{F1T} \\
F_{2}^{(+)} & = \frac{2}{B} \mathcal{F}_{-+}^*(z_C,z_T) \mathcal{F}_{+-}(z_C,z_T) , \label{F2p} \\
F_{2}^{(-)} & = \frac{2}{B} \mathcal{F}_{--}(z_C,z_T) \mathcal{F}_{++}^*(z_C,z_T) , \label{F2m}
\end{align}
It is worth observing that, while the spatial frequencies are fixed only by the combinations $k d_T/f_T$ and $k d_C/f_C$, the background $B$ and the Fourier coefficients depend on all the parameters of the setup, except the mask-to-detector distances. Therefore, determining these coefficients is the key to estimate one of the longitudinal distances or one of the mask center positions. Notice that, in the main text, the correlation function has been conveniently redefined dividing by the background $B$.

In the paper, we consider the case in which the double slits axes coincides with the respective optical axes ($X_C=X_T=0$) and the source is characterized by a Gaussian intensity profile, of width $\sigma$:
\begin{equation}\label{sourcegauss}
\mathcal{S}(x_s) = \mathcal{S}_0 \exp \left( - \frac{x_s^2}{2\sigma^2} \right) ,
\end{equation}
for which
\begin{equation}\label{Phipeval}
\Phi(y) \propto \exp\left\{ - \frac{y^2}{2 \left[ \frac{1}{(k\sigma)^2} + \sigma^2\left( \frac{1}{z_T}-\frac{1}{z_C} \right)^2 \right] } \left[ 1 + \ii k \sigma^2 \left( \frac{1}{z_T}-\frac{1}{z_C} \right) \right] \right\} .
\end{equation}
In this case, the Fourier coefficients read 
\begin{equation}
F_{1C}=F_{1T}=  \frac{\cos(\alpha\beta/2)}{\cosh(\alpha/2)}, \quad F_2^{(+)} = \frac{1}{1+\exp(\alpha)} = 1- F_2^{(-)}
\end{equation}
with
\begin{equation}
\alpha= \frac{1}{1+\beta^2}\frac{d_C}{\sigma_C}\frac{d_T}{\sigma_T}, \quad \beta= \sigma\left|\frac{1}{\sigma_T}-\frac{1}{\sigma_C}\right|
\end{equation}
as reported in the main text.

\section{Effects of turbulence surrounding the masks}

Turbulence around one or both masks is generally detrimental for the correlation measurement. The effect can be modelled by the replacing $A_J(x_o)$, with $J=C,T$, as defined in Eq.~(\ref{eq:AJ}), with
\begin{equation}
A_J(x_o) = a \left[ e^{\ii \phi_J^{(-)}} \delta \left(x_o+\frac{d_J}{2}\right) + e^{\ii \phi_J^{(+)}} \delta \left(x_o-\frac{d_J}{2}\right) \right] .
\end{equation}
where $\phi_J^{(\pm)}$ are four random phases, affecting the field propagating through each slit $q=\pm$ to each detector $\mathrm{D}_J$ with $J=C,T$. These phases are generally not correlated with each other. The effect of the presence of turbulence consists in
\begin{equation}
\mathcal{F}_{qq'}(z_C,z_T) \longrightarrow \exp\left( \ii (\phi_T^{(q)} -  \phi_C^{(q')} ) \right) \mathcal{F}_{qq'}(z_C,z_T).
\end{equation}
Therefore, if the four phases $\phi_J^{(\pm)}$ are not correlated with each other, turbulence in the setup illustrated in Fig.~1(a)-(b) of the main text affects the final result \eqref{Gammafourier}, defined by the turbulence-sensitive coefficients \eqref{B}--\eqref{F2m}.

The situation is different for the setup in Fig.~1(c), since there, as the two masks physically coincide, 
\begin{equation}\label{Atilde}
d_C=d_T=d, \quad z_C=z_T=z, \quad \phi_C^{(\pm)} = \phi_T^{(\pm)} = \phi^{(\pm)} .
\end{equation}
Therefore, the functions $\mathcal{F}_{++}$ and $\mathcal{F}_{--}$ are both insensitive to turbulence, as well as the coefficient $F_2^{(-)}$ in which they appear. In the case where full second order correlation interference occurs, namely when only the turbulence-free coefficient $F_2^{(-)}$ is nonvanishing, the correlation function $\Gamma(x_C,x_T)$ can be considered robust with respect to turbulence effects modelled as in Eq.~\eqref{Atilde}.

\twocolumngrid

\end{document}